\documentclass{elsarticle}
\usepackage[utf8]{inputenc}
\usepackage{amsmath,amssymb,mathrsfs}

\newtheorem{theorem}{Theorem}

\newtheorem{remark}[theorem]{Remark}

\title{Parameter Estimation and System Identification for Continuously-Observed Quantum Systems}

\author[1]{Hendra I. Nurdin\corref{cor1}}
\ead{h.nurdin@unsw.edu.au}

\author[2]{M\u{a}d\u{a}lin Gu\c{t}\v{a}}
\ead{Madalin.Guta@nottingham.ac.uk}

\address[1]{School of Electrical Engineering and Telecommunications, UNSW Australia, Sydney NSW 2052, Australia.}
\address[2]{School of Mathematical Sciences, University of Nottingham, University Park
Nottingham, NG7 2RD, UK}
 
\begin{document}

\begin{abstract}
This paper gives an overview of parameter estimation and system identification for quantum input-output systems 
by continuous observation of the output field. We present recent results on the quantum Fisher information of the output with respect to unknown dynamical parameters. 
We discuss the structure of continuous-time measurements as solutions of the quantum Zakai equation, and their relationship to parameter estimation methods. 
Proceeding beyond parameter estimation, the paper also gives an overview of the emerging topic of quantum system identification for black-box modeling of quantum systems by continuous observation of a traveling wave probe, for the case of ergodic quantum input-output systems and linear quantum systems. Empirical methods for such black-box modeling are also discussed. 
\end{abstract}

\maketitle

\section{Introduction}

Quantum input-output (I/O) dynamics is an effective framework for modelling the evolution of Markovian quantum open systems by coupling with a traveling quantum field such as a coherent laser beam \cite{GZ04,WM10,CKS17}.
After the interaction between the system and the field, certain field observables can be continuously observed by a quantum non-demolition (QND) measurement (see \cite{BVT80,Belavkin94,Lupascu07} and the references therein) producing a classical stochastic process or a quantum trajectory \cite{Carmichael,DCM,WM93} as measurement record. The measurement record from a continuous QND measurement of a traveling quantum field  can be used to extract information about the system via the process of quantum filtering \cite{Belavkin94,BvHJ07}, the quantum analogue of stochastic filtering for classical stochastic Markov processes.

To our knowledge, the earliest treatment of parameter estimation on  \\ continuously-observed quantum systems appeared in \cite{Mab96}. This work considers estimating the vacuum Rabi frequency $g$ that characterises the strength of the coherent coupling between an atom and a cavity mode, while a photon counting measurement is performed on the output of the cavity. The paper used a Bayesian approach where the posterior distribution of the unknown parameter was computed using the likelihood function of the observed quantum trajectory.
Many other works have since  addressed related problems of parameter estimation based on stochastic records obtained via probe measurements, see for instance \cite{GW01,CG09,GM13,KM16} and the references therein. 

A problem related to our setting is that of the estimation of a classical signal (which may be interpreted as a time-dependent parameter) coupled to a quantum I/O system \cite{VDM01,TWC11,Tsang09,CG09}. Reference \cite{VDM01} considers a classical signal that modulates the position of a quantum harmonic oscillator while this is continuously-observed  by coupling it to a probe. In \cite{TWC11}, a quantum Cram\'{e}r-Rao bound was derived for the estimation of a signal based on the continuous-measurement of a quantum sensor. In the case of the estimation of a signal that modulates the position of a quantum harmonic oscillator in which all signals and noises involved are stationary, it was shown that the bound can be saturated by using a combination of coherent noise cancellation and  time-symmetric quantum smoothing  \cite{Tsang09} on the sensor output. In the present paper, the primary focus will be on continuously-observed systems with \emph{fixed} unknown parameters and the estimation of those parameters, and we will not consider the estimation of a classical signal through a quantum sensor. 

System identification is concerned with black-box modelling of an unknown dynamical system from externally observed input and output signals, without a priori knowledge of the system's internal structure. It is an important and  well-established topic in the classical (non-quantum) setting \cite{Ljung99} and is closely related to the subject of time series modelling \cite{FY03}. In system identification, one chooses a model to fit to the unknown system and then designs appropriate inputs aimed at maximising the information that can be gained from observing the system's output responses to these inputs. Based on the collected input-output data, the model parameters are estimated and the resulting model is  tested through a model validation phase; further details can be found in section \ref{sec:qsys-id}.   

In the quantum setting, system identification for black-box closed Hamiltonian systems was proposed in \cite{BY12},  based on repeated projective measurements to estimate the quantum expectations of certain system observables. These estimated expectation values are used as the data with which to fit a Hamiltonian to the unknown closed system. A similar approach has been pursued for open quantum systems, e.g., \cite{ZS15,SC17} and the references therein, using time traces of estimated quantum expectation values. On the other hand, quantum system identification for  continuously observed quantum systems was initiated  in \cite{GY16} with the study of the identification of single input single output (SISO) \emph{passive} linear quantum I/O  systems \cite{NY17}, and continued in \cite{LG17} with the case of general linear quantum systems.
In this setting, the \emph{transfer function} encapsulates all the information that can be captured by observing the system's output response for known time-dependent inputs;
the papers show that for these two classes the system matrices can be identified up to unitary and respectively symplectic transformations, extending similar results for classical linear systems.

The identifiability of linear quantum systems driven by \emph{stationary} (time-independent) Gaussian quantum noise (see \cite[\S II-E]{GJN10} for an overview)  was investigated in  \cite{LG17,LGN18}. In this case the information is captured by the \emph{power spectrum} of the output. For pure input Gaussian states, two \emph{globally minimal} systems have the same power spectrum if and only if they have the same transfer function, and are therefore related by a symplectic transformation \cite{LGN18}. 

The contribution \cite{GK17} develops the system identification and \emph{information geometry} theory for finite, non-linear ergodic quantum I/O systems. Similarly to the linear case, in the stationary regime, output-equivalent systems are related by a certain group of transformations acting on the hamiltonian and jump operators. 
On the statistical side, it is shown that the output quantum Fisher information (QFI) for unknown, identifiable parameters grows linearly in time and the explicit rate is computed as a quantum Markov covariance of certain generators governing parameter changes. An alternative QFI formula has been obtained in \cite{Molmer2014}. Both approaches are intimately connected to the quantum trajectories approach to dynamical phase transitions (DPT) for open quantum systems \cite{GarrahanLesanovsky}, and indicate that systems near a DPT may exhibit large QFI, with potential application to quantum metrology \cite{GutaMacieszczakGarrahanLesanovsky}.

Devising continuous-time measurements that achieve the QFI rate is currently one of the main open problems in this area. In contrast, the statistical performance of homodyne and counting measurements has been investigated in  \cite{GM13,KM16,GutaCatanaBouten,GutaCatanaKypraios} but a general theoretical understanding is still lacking. 

The results in \cite{GY16,GK17,LG17,LGN18} contribute to the theoretical  foundation  of system identification of quantum I/O systems. However, they do not provide methods and algorithms to identify a black-box model  given empirical single measurement records of the output. Such algorithms are crucial for the practical use  of system identification and this aspect is discussed in section \ref{subsec:lqs-id}. 

This paper is structured as follows. Section \ref{sec:model} introduces the class of quantum I/O systems and their modeling by quantum stochastic differential equations.  Section \ref{sec:q-meas-filt} reviews the notion of quantum filtering, the quantum Zakai equation and positive operator valued measures (POVMs) for continuously-observed quantum systems from solutions of the quantum Zakai equation. This is followed in section \ref{sec:estimation} with an overview of the quantum parameter estimation problem, the quantum Fisher information for parameter estimation of quantum I/O systems, and estimation methods for `standard' continuous-time measurements such as counting and homodyne detection. Section \ref{sec:qsys-id} introduces the quantum system identification problem for ergodic quantum I/O systems and linear quantum systems and discusses empirical methods for black-box system identification of linear quantum systems. Finally, section \ref{sec:conclu} discusses open problems and directions for future research.

\noindent \textbf{Notation.} For the remainder of the paper, we will use the following notation. $X^{\top}$ denotes the transpose of a matrix $X$, $X^{\dag}$ denotes the adjoint of a Hilbert space operator $X$ and if $X=[X_{jk}]$ is a matrix of operators then $X^{\dag}$ is the conjugate transpose of $X$, $X^{\dag} =[X_{kj}^{\dag}]$. $I_n$ will denote an $n \times n$ identity matrix and $I$ can denote either an identity matrix (whose dimension can be inferred from the context), an identity map or an identity operator. $\mathrm{Tr}$ denotes the trace of a matrix or an operator and  $\mathrm{Im}(X)$ denote the elementwise real part of a matrix $X$. For a signal (a function of time) $Y$, $Y_{0:t}=\{Y_{\tau}\}_{0 \leq \tau \leq t}$. 

\section{Mathematical model}
\label{sec:model}
In this review, to focus on the main ideas we consider only the case of a quantum I/O system  (equivalently, a quantum Markov model) that is coupled only to a single traveling field. We consider a one-dimensional field on the $x$-axis travelling from right to left and the quantum system is located at $x=0$. 
We shall refer to the quantum system that is coupled to the field as the principal quantum system. Under some physical assumptions and approximations, in a large class of physical scenarios of interest the unitary propagator $U_t$ on the system and the field is given by a  Hudson-Parthasarathy quantum stochastic differential equation (QSDE) \cite{HP84}:
\begin{align}
dU_t &= (-(iH+(1/2)L^{\dag}L)dt + dB^{\dag}_t L - L^{\dag}S dB_t + (S-I) d\Lambda_t)U_t,\,U_0=I. \label{eq:HP-QSDE}  \end{align}
Here $B_t$, $B^{\dag}_t$ and $\Lambda_t$ are the annihilation, creation and gauge process of the traveling field, $H$ is the principal system Hamiltonian, $L$ is the coupling operator of the principal to the field creation operator, and $S$ is a unitary matrix ($S^{\dag}S=S^{\dag}S=I$) representing the coupling of the system to the gauge process of the field. The three processes $B_t$, $B^{\dag}_t$ and $\Lambda_t$ are referred to as {\em fundamental processes}. For a review of this class of models, we refer to \cite{BvHJ07,GJ09}.

The time evolution of a principal system operator $X$, in the Heisenberg picture with respect to the propagator \eqref{eq:HP-QSDE} is given by $j_t(X)$, where $j_t(X)=U_t^{\dag} X U_t$. It is given  by the QSDE:
\begin{align}
dj_t(X) &=\mathcal{L}_{j_t(L),j_t(H)}(j_t(X)) dt + dB_{t}^{\dag}j_t(S)[j_t(X),j_t(L)] + \notag \\
&\qquad [j_t(L^{\dag}),j_t(X)]dB_{t}+\mathrm{tr}(j_t(S^{\dag})j_t(X) j_t(S)-j_t(X))d\Lambda_{t}, \label{eq:X-evolution} 
\end{align}
where $\mathcal{L}_{Y,Z}(X)$ is a map defined by:
$$
\mathcal{L}_{Y,Z}(X) = i[Z,X] + (1/2) \left( Y^{\dag}[X,Y] + [Y^{\dag},X]Y  \right).
$$

Due to the interaction with the system, the fundamental processes that impinges upon the system at time $t$, considered as an input to the system, undergoes an instantaneous transformation according to $M_{{\rm o},t}=U_t^{\dag} M_t U_t$, where $M_t$ can be any of the fundamental processes or linear combinations thereof, producing output fields. Let $W^{Q}_t = B_{t} + B^{\dag}_t$ and   $W^{P}_t = -iB_t + i B^{\dag}_t$ be the amplitude and phase quadratures of $B_t$, respectively. Then $W^{Q}_t$, $W^{P}_t$ and $\Lambda_t$ undergo an instantaneous transformation after interaction with the principal to become the  output field processes $W^{Q}_{{o},t}$, $W^{P}_{{o},t}$ and $\Lambda_{{ o},t}$ given by the QSDE: 

\begin{align*}
dW^{Q}_{{ o},t} &=j_t(L+ L^{\dag})dt + j_t(S) dB_{t}  +  j_t(S^{\dag})dB^{\dag}_{t} \\
dW^{P}_{{o},t} &=j_t(-iL+ i L^{\dag})dt -i j_t(S) dB_{t}  + i  j_t(S^{\dag})dB^{\dag}_{t} \\
d\Lambda_{{o},t}  &=  j_t(L^{\dag})j_t(L) dt  + j_t(S^{\dag})j_t(L) dA_{t}^{\dag}+ j_t(L^{\dag})j_t(S) dA_{t} +d\Lambda_{t}.     
\end{align*}

Two crucial properties of quantum I/O models are:
\begin{enumerate}
\item $[M_{{o},t},M_{{o},s}]=0$ for all $s,t \geq 0$ when $M_{{ o}}$ is any of $W^{Q}_{{o}}$, $W^{P}_{{ o}}$ and $\Lambda_{{o}}$,  known as the {\em self-non-demolition} property. It follows that they can be mapped to classical stochastic processes and the measurement of these processes is a QND measurement. 

\item  $[j_t(X),M_{{o},s}]=0$ for all principal system operators $X$ and all $0\leq s \leq t$ when $M_{{o}}$ is any of $W^{Q}_{{ o}}$, $W^{ P}_{{ o}}$ and $\Lambda_{{ o}}$, known as the {\em non-demolition property}.  It implies that the quantum conditional expectation of $j_t(X)$ onto $M_{{ o},0:t}$ exists. 
\end{enumerate}

Measurements of $W^{Q}_{{o}}$, $W^{P}_{{ o}}$ are often referred to as diffusive measurements, while measurement of $\Lambda_{{o}}$ is referred to as a counting measurement. It is common and often useful to consider the Schr\"{o}dinger picture in which the system-field state evolves in time by applying the unitary $U_t$ to the initial state. This gives the state 
$$
\tau_t = U_t(\rho_0 \otimes |0_{\rm f}\rangle\langle 0_{\rm f}|) U_t^\dagger
$$
where $\rho_0$ is the initial state of the principal system while $|0_{\rm f}\rangle$ is the vacuum state of the field. The reduced system state is $\rho_t= {\rm Tr}_{\mathcal{H}_f} (\tau_t)$ and satisfies the Lindblad equation
$$
\dot{\rho}_t = \mathcal{L}^{\star}_{H,L}(\rho_t)
$$
where $\mathcal{L}^{\star}_{H,L}$ is the dual map to $\mathcal{L}_{H,L}$ defined as 
\begin{equation}
\mathcal{L}_{H,L}^{\star}(X) = i[X,H] +L XL^{\dag} -(1/2)L^{\dag}L X -(1/2) X L^{\dag}L. \label{eq:Lindblad-dual}
\end{equation}
If the principal system is finite dimensional, we call the I/O dynamics ergodic if there exists a unique full rank stationary state $\rho_{ss}$ (such that $\mathcal{L}^{\star} (\rho_{ss}) =0)$. 
In this case we have convergence to stationarity
$$
\lim_{t\to\infty}\rho_t = \rho_{ss}
$$
and the convergence takes place exponentially on a time scale of the order of the inverse of the spectral gap of  $\mathcal{L}^{\star}$.

\section{Quantum filtering, quantum Zakai equation and POVMs for \\ continuously-observed quantum systems}
\label{sec:q-meas-filt}
\subsection{Quantum filtering equation}
Let the principal system be prepared in the state $\rho_0$. The joint initial state of the system and field is then $\rho_{\rm pf}= \rho_0 \otimes |0_{\rm f}\rangle \langle 0_{\rm f}|$.  Let $\mu_{\rho_{\rm pf}}(\cdot)=\mathrm{tr}(\rho_{\rm pf}\cdot)$ be a state (i.e., the quantum expectation operator) and $\mu_{\rho_{\rm pf}}(j_t(X) \mid M_{{\rm o},0:t})$ denote the quantum conditional expectation of $j_t(X)$ onto $M_{{\rm o},0:t}$ in the state $\mu_{\rm pf}$ \cite{BvHJ07}. We also introduce the shorthand notation $\pi_t(X)=\mu_{\rho_{\rm pf}}(j_t(X) \mid M_{{\rm o},0:t})$. Depending on the continuous observation made, $\pi_t(X)$ will be given by given by a QSDE. For example, under a continuous QND measurement of $W^{Q}_{{o},\cdot}$ the quantum filtering equation takes the form:
\begin{eqnarray*}
d\pi_t(X) &= \pi_t(\mathcal{L}_{L,H}(X))dt +\biggl(\pi_t(X L + LX)- \pi_t(L+L^{\dag})\pi_t(X)\biggr)dI_t, 
\end{eqnarray*}
where $I_t$ is a quantum innovation process given by
$$
I_t =  W^{Q}_{{ o},t} -\int_{0}^t  \pi_{\tau}(L + L^{\dag})d\tau,
$$
or, in differential form,
$$
dI_t = dW^{\rm Q}_{{\rm o},t} - \pi_t(L + L^{\dag})dt,\; I_0=0. 
$$
In the case of measurement of $\Lambda_{0,t}$ (photon counting) then the SME takes the form:
$$
d\pi_t(X) = \pi_t(\mathcal{L}_{L,H}(X))dt +  \left(\frac{\pi_t(L^{\dag}XL)}{\pi_t(L^{\dag}L)} -\pi_t(X)\right)dI_t, 
$$
where
$$
I_t = \Lambda_{o,t}-\int_{0}^t \pi_{\tau}(L^{\dag}L)d\tau.
$$

The quantum filtering equation is an operator-valued equation since all processes are operator-valued. However, the processes are self-commuting (at different times) and commuting with each other, so they can be treated as classical stochastic processes (and can be mapped to such). With this in mind we can write 
$$
\pi_t(X) = \mathrm{tr}(\rho_{c,t} X),
$$
where $\rho_{c,t}$ is a stochastic density operator satisfying a stochastic differential equation (SDE) known as a stochastic master equation (SME). In the case of continuous observation  of $W_{o}^{\rm Q}$ the SME takes the form of the density operator-valued SDE:
\begin{equation}
    \label{eq:state.filter.h}
d\rho_{c,t} = \mathcal{L}^{\star}_{L,H}(\rho_{c,t})dt +  (L\rho_{c,t} + \rho_{c,t} L^{\dag} -\mathrm{Tr}((L+L^{\dag})\rho_{c,t})\rho_{c,t})dI_t, 
\end{equation}
where $\mathcal{L}_{L,H}^{\star}$ is the map given by \eqref{eq:Lindblad-dual}. For a measurement of $\Lambda_{o,\cdot}$ the SME takes the form:
\begin{equation}
    \label{eq:state.filter.c}
d\rho_{c,t} = \mathcal{L}^{\star}_{L,H}(\rho_{c,t})dt +  \left(\frac{L \rho_{c,t} L^{\dag}}{\mathrm{Tr}(\rho_{c,t}L^{\dag}L)} -\rho_{c,t} \right) dI_t.
\end{equation}

\subsection{Quantum Zakai equation, POVMs and likelihood functions for continuously- observed systems}
\label{sec:Zakai-likelihood}

The conditional expectation $\pi_t$ can be expressed as the ratio:
$$
\pi_t(X) = \sigma_t(X) \sigma_t(I)^{-1},
$$
where $\sigma_t(X)$ and  $\sigma_t(I)$ are two commuting processes satisfying a linear QSDE known as the quantum Zakai equation. In the case of the measurement of $Y = W_{{\rm o}}^{\rm Q}$, the Zakai equation takes the form:
\begin{equation}
d\sigma_t(X) = \sigma_t( \mathcal{L}_{L,H}(X))dt + (\sigma_t(XL + L^{\dag} X)dY_t, \label{eq:Zakai-diffusive}
\end{equation}
for any operator $X$ on the principal system.  The initial condition for the equation is given by $\sigma_0(X)= {\rm tr}(\rho_0 X)$,  where $\rho_0$ is the initial state of principal system.  Writing $\sigma_t(X) ={\rm tr}(\varrho_t X)$, the unnormalised density matrix $\varrho_t$ satisfies :
$$
d\varrho_t  = \mathcal{L}_{L,H}^{\star}(\varrho_t)dt + (L \varrho_t + \varrho_t L^{\dag})dY(t)
$$

For a photon counting measurement $Y =\Lambda_{o}$ the Zakai equation takes the form,
\begin{equation}
d\sigma_t(X) = \sigma_t( \mathcal{L}_{L,H}(X))dt + (\sigma_t(L^{\dag} X L)-\sigma_t (X))(dY_t-dt), \label{eq:Zakai-counting}
\end{equation}
and the unnormalized density matrix has takes the form,
$$
d\varrho_t  = \mathcal{L}_{L,H}^{\star}(\varrho_t)dt + \left(L \varrho_t L^\dagger- \varrho_t\right)(dY_t-dt)  
$$

The equation for the unnormalized density operator has the explicit solution:
\begin{equation}
\varrho(t) = \overleftarrow{T} e^{ \mathscr{L}_{t} } \rho_0, \label{eq:Zakai-exp}
\end{equation}
where  $\mathscr{L}_t$ is a superoperator given by the stochastic integral
$$
\mathscr{L}_t (\varrho) =  \mathcal{L}_{L,H}^{\star}(\varrho)t + \int_{0}^t (L \varrho  + \varrho  L^{\dag})dY_{\tau}
$$
or
$$
\mathscr{L}_t (\varrho) =  \mathcal{L}_{L,H}^{\star}(\varrho)t + \int_{0}^t \left( L \varrho L^\dagger- \varrho \right)(dY_\tau-d\tau)
$$
in the case of the measurement of $W_{o}^Q$ and $\Lambda_{o}$, respectively. 

The stochastic time-ordered exponential
\begin{equation}
\Phi_t(Y_{0:t})= \overleftarrow{T} e^{ \mathscr{L}_t  }   
\end{equation}
in the solution of the Zakai equation \eqref{eq:Zakai-exp} is associated with a positive operator-valued measure (POVM) for continuous measurements. For a fixed time $T$, let $\mathscr{Y}_T$ denote the $\sigma$-algebra generated by the observation $Y_{0:T}$ (viewing $Y$ as an equivalent stochastic process). For any initial state $\rho$ of the system, the $\mathscr{Y}_T$-measurable function 
$$
\mathrm{tr}(\Phi_T(Y_{0:T}) \rho)
$$ 
is the Radon-Nikodym derivative of the underlying probability measure on $\mathscr{Y}_T$  with respect to an appropriate reference measure \cite{BG13}. In the case of the diffusive measurement $Y=W_{o}^Q$  this reference measure is the Wiener measure on $\mathscr{Y}_T$ while for a photon counting measurement $Y=\Lambda_{o}$ this reference measure is the Poisson measure with intensity 1. 
Therefore, we can define the POVM $\Pi$ associated with the measurement of $Y$ as 
$$
\Pi(A)= \int_{A}  \Phi_T(\omega)\mu(d\omega)
$$
for any $A \in \mathscr{Y}_T$, where $\mu$ is the reference measure on $\mathscr{Y}_T$. Therefore, $\mathrm{tr}(\Pi(A)\rho)$ for any initial density operator gives the probability of observing a trajectory (over the time interval $[0,T]$) that lies in $A$, when $Y$ is continuously observed over the interval $[0,T]$ and the system is initialized in the state $\rho$. 

Heuristically, as a Radon-Nikodym derivative, $\mathrm{tr}(\Phi_T(Y_{0:T}) \rho)$ may be viewed as 
a ``probability density function" with respect to the underlying reference measure. For a single trajectory  $y_{0:T}$ as a realization of $Y_{0:T}$, it follows from the discussion above that the function
\begin{equation}
\ell(y_{0:T})= \mathrm{tr}(\Phi_T(y_{0:T})\rho) =\mathrm{tr}(\sigma_T(I)) \label{eq:likelihood}   
\end{equation}
is the likelihood function of the trajectory. Such a likelihood function is the basis of the maximum likelihood approach to quantum parameter estimation and system identification that will be discussed later on. 

From the expression for the likelihood, it can be straightforwardly shown that the log likelihood $\log \ell(Y_{0:T})$ satisfies the equations
\begin{equation}
    \label{eq:log.lik.h}
d\log\ell(Y_{0:t}) = 
{\rm tr} (L \rho_{c,t} + \rho_{c,t}L^{\dagger} )(dY_t - {\rm tr} (L\rho_{c,t} +\rho_{c,t} L^{\dagger} ) dt)
\end{equation}
or
\begin{equation}
    \label{eq:log.lik.c}
d\log\ell(Y_{0:t}) = (1 - {\rm tr} (L^\dagger L \rho_{c,t}))dt + dY_t \ln ({\rm Tr} (L^\dagger L \rho_{c,t}))
\end{equation}
for a measurement of $W^Q_o$ and $\Lambda_o$, respectively.

\begin{remark}
We note that \cite{GM13} gives a heuristic derivation of the Zakai equation for the photon counting case when the reference measure is a Poisson measure with intensity $\lambda>0$ not necessarily equal to unity. The equations in this case become
$$
d \varrho_t = \mathcal{L}^{\star}_{L,H}(\varrho_t)dt  + 
\left( \frac{L \varrho_t L^\dagger }{\lambda}- \varrho_t\right)(d\Lambda_{o,t}-\lambda dt) .
$$

$$
d\log\ell(Y_{0:t}) = (\lambda - {\rm tr} (L^\dagger L \rho_{c,t}))dt + d\Lambda_{o,t} \log ({\rm Tr} (L^\dagger L \rho_{c,t}) /\lambda)
$$
\end{remark}

\section{Quantum parameter estimation}
\label{sec:estimation}
In quantum parameter estimation we consider the scenario of a quantum I/O system that has dependence on a vector of $k$ {\em unknown} parameters $\theta=(\theta_1,\theta_2,\ldots,\theta_k)$ through either the Hamiltonian $H_\theta$ or coupling operator $L_\theta$ or both of them. For instance, if $L$ is fixed and known and $H_\theta=\sum_{i=1}^k \theta_i H_i$ then estimating $\theta$ amounts to a Hamiltonian identification problem. Below we give an overview of the general quantum parameter estimation theory, and describe how this applies to the case of parameter estimation on quantum I/O systems. 

\subsection{Quantum Cramer-Rao lower  bound and quantum Fisher information}

Consider a quantum system whose state $\rho^{\theta}$ depends smoothly on an \emph{unknown} multidimensional parameter $\theta\in \mathbb{R}^k$. To estimate $\theta$, we perform a measurement and construct an estimator $\hat{\theta} = f(X)$ where $X$ is a vector of measurement outcomes. According to the quantum Cram\'{e}r-Rao (QCR) bound \cite{Holevo,BraunsteinCaves}, 
the covariance matrix of any \emph{unbiased} estimator $\hat{\theta}$ is lower bounded as
\begin{equation}
    \label{eq:QCRB}
\mathrm{Cov} (\hat{\theta}):= 
\mathbb{E}\left[ (\hat{\theta} - \theta)(\hat{\theta} - \theta)^t\right] \geq 
F(\theta)^{-1}  
\end{equation}
where the right side is the inverse of the quantum Fisher information (QFI) matrix, which is defined as
$$
F(\theta)_{ij} = \frac{1}{2}\mathrm{Tr}\left(\rho^{\theta} (S^{\theta}_iS^{\theta}_j + S^{\theta}_jS^{\theta}_i)\right),
$$
with $S^{\theta}_1,\dots ,S^{\theta}_k$ the \emph{symmetric logarithmic derivatives} defined via the Lyapunov equation
$$
\rho^\theta_i =\frac{1}{2}(S^{\theta}_i\rho^{\theta} + \rho^{\theta} S^{\theta}_i), \qquad {\rm where~~}\rho^\theta_i:=\frac{\partial \rho^{\theta}}{\partial \theta_i}.
$$
In particular, if $\rho^\theta$ is a family of pure states of the form 
$\rho^\theta = |\psi^\theta\rangle \langle \psi^\theta| $ with $|\psi^\theta\rangle= e^{-i\theta G} |\psi \rangle$ for some given reference state $|\psi \rangle$ and selfadjoint generator $G$, then the QFI is independent of $\theta$ and is proportional to the variance of $G$
\begin{equation}
\label{eq:QFI.pure}
F= 4 {\rm Var}_\psi (G) = 
4 
(\langle \psi | G^2 |\psi\rangle - 
\langle \psi | G |\psi\rangle^2). 
\end{equation}

In general, the QCR bound \eqref{eq:QCRB}  is not achievable for a single quantum system, but it is \emph{asymptotically} achievable for \emph{one-dimensional parameters} in the limit of large sample size $n$, i.e. for independent systems with joint state $\left(\rho^\theta\right)^{\otimes n}$. 
In this case, by measuring $S^{\theta}$ (more precisely we measure $S^{\tilde{\theta}}$ for some rough estimate $\tilde{\theta}$ obtained from a small subsample \cite{GillMassar}), and computing the maximum likelihood estimator $\hat{\theta}_{\rm ML}$ (see section \ref{sec:ML} for the definition) one obtains 
$$
\lim_{n\to\infty}
n\, \mathbb{E}\left[(\hat{\theta}_{\rm ML} - \theta)^2 \right]\to F(\theta)^{-1}.
$$
In addition, under appropriate regularity conditions, $\hat{\theta}_{\rm ML}$ has an asymptotically normal distribution with variance $F(\theta)^{-1}$. For higher dimensional parameters, the QCR bound on the covariance matrix is asymptotically achievable if and only if ${\rm Im}( F(\theta)) =0$. When this is not the case, one aims to replace the QCR matrix lower bound, by a bound for the mean square error ${\rm MSE} (\hat{\theta}):= {\rm Tr} ({\rm Cov} (\hat{\theta}))$ of the estimator $\hat{\theta}$, or other quadratic forms of the covariance. The trivial bound ${\rm Tr} (F(\theta)^{-1})$ follows from the QCR bound \eqref{eq:QCRB} but is generally not achievable. A more refined bound was introduced by Holevo \cite{Holevo}
\begin{equation}
\label{HCR}
{\rm MSE} (\hat{\theta})
\geq H(\theta):=
\min_{\bold{X},V}\left\{
{\rm Tr}(V)\, : \, V\geq {\rm Tr}(\rho^\theta \mathbf{X} \mathbf{X}^T)
\right\},
\end{equation}
where the minimum runs over all  $k\times k$ \emph{real} matrices $V$, and $k$-tuples of selfadjoint system operators $X = (X_1,\dots,X_k)^T$,  which satisfy the constraints 
${\rm Tr}(\rho^\theta_i X_j) =\delta_{i,j}$ for all $i,j=1,\dots ,k$. 
The Holevo bound is at most twice as large as the simple bound ${\rm Tr} (F(\theta)^{-1})$ \cite{AlbarelliTsangDatta}; using the theory of local asymptotic normality \cite{KahnGuta} it can be shown that the Holevo bound is asymptotically achievable \cite{GutaKahn06,DemkowiczGuta} (more precisely, it is equal to the minimax constant of the asymptotic estimation problem). We conclude that, in spite of its limitations, the QFI is a key tool  in assessing the limits of precision in quantum 
estimation and we will return to it when analysing the statistical structure of quantum Markov models.

\subsection{Estimation of quantum I/O dynamics}

We now consider the problem of estimating dynamical parameters of quantum I/O systems by means of output measurements. Let us assume that the Hamiltonian and jump operators depend on an unknown parameter $\theta\in \mathbb{R}^k$ so that $H= H_\theta, L= L_\theta$, and for simplicity we take $S=I$. Therefore the unitary evolution depends on $\theta$ and we denote by $U^\theta_t$ the corresponding unitary, cf. equation \eqref{eq:HP-QSDE}. Furthermore we assume that the dynamics is ergodic and denote the unique stationary state by $\rho^\theta_{ss}$. If the initial system state is 
$|\psi_0\rangle$ then the system-output state at time $t$ is given by the vector
$$
|\psi^\theta_{so,t}\rangle = U^\theta_t (|\psi_0\rangle \otimes |0_{\rm f}\rangle). 
$$
This state exhibits finite-time correlations of the order of the convergence time to stationarity, and can be seen as a continuous-time generalization of a matrix product state \cite{VerstraeteCirac}.
The output state can be written as 
$$
\rho^{\theta}_{o,t} = {\rm Tr}_{\mathcal{H}_s} 
\left[U^\theta_t(\rho_0 \otimes |0_{\rm f}\rangle \langle 0_{\rm f}|) U^{\theta \dagger}_t \right], \qquad \rho_0 = |\psi_0\rangle\langle \psi_0|
$$
where $\mathcal{H}_s$ is the principal system Hilbert space. 
The QFI of the states $|\psi^\theta_{so,t}\rangle$ and $\rho^{\theta}_{o,t}$ has been investigated in \cite{Molmer2014,GK17,GutaCatanaBouten}, while the discrete time case has been analysed in \cite{Guta2011,GutaKiukas1}. While the former state is generally more informative than the latter, for large $t$ the QFI of both states grows linearly in time with QFI rate
\begin{equation}\label{eq:qfi.markov}
F(\theta)_{ab} = 
\mathrm{Tr}
\left[\rho^\theta_{ss}  
\left(\dot{L}_{\theta,a} - i[L_\theta , \mathcal{L}_\theta^{-1}(\dot{E}_{\theta,a})  ]\right)^\dagger\cdot
\left(\dot{L}_{\theta,b} - i[L_\theta , \mathcal{L}_\theta^{-1}(\dot{E}_{\theta,b})  ]\right) \right]
\end{equation}
where 
$$
\dot{E}_{\theta,a} :=  \dot{H}_{\theta,a} +  {\rm Im} (\dot{L}^{\dagger}_{\theta,a} L_\theta  )  - 
{\rm Tr} \left[\rho_{ss}^\theta (\dot{H}_\theta +  {\rm Im} (\dot{L}^{\dagger}_{\theta,a}L_\theta  ) ) \right] I
$$
and $\mathcal{L}_\theta^{-1}$ denotes the inverse of the restriction of $\mathcal{L}_\theta$ to the space of zero-mean operators $\{A: {\rm Tr}( \rho_{ss}^\theta A)= 0\}$. An alternative QFI formula can be found in \cite{Molmer2014}, expressed in terms of the dominant eigenvalue of a deformed Lindblad operator.

Note that the expression \eqref{eq:qfi.markov} is explicitly positive and its magnitude is relatedt to the spectral gap of the generator $\mathcal{L}_\theta$, so that systems with small gap may exhibit large QFI, with potential applications for quantum enhanced metrology \cite{GutaMacieszczakGarrahanLesanovsky}. For example, consider the simple case of a system with fixed Hamiltonian $H$ and jump operator $L_\theta = e^{-i\theta}L$; in this case, the system-output state has the following dependence on $\theta$
$$
|\psi^\theta_{so,t}\rangle= 
e^{-i\theta \Lambda_{t}}
|\psi^0_{so,t}\rangle.
$$
where $\Lambda_{t}$ is the counting operator. This can be understood in terms of the unraveling of $|\psi_{so,t}\rangle$ as superposition of $p$-photon quantum 
trajectories of the form
$$
e^{-i H_{e} (t-t_p)} \cdot L\cdot e^{-i H_{e} (t_p-t_{p-1})} \cdot \dots \cdot L \cdot e^{-i H_{e} t_1}|\psi_0\rangle \otimes|t_1,\dots t_p\rangle
$$
where $|t_1,\dots , t_p\rangle$  is the $p$-photon (singular) field state with excitations at times $(t_1,\dots ,t_p)$, and $H_{e} = H-iL^\dagger L /2$ is the \emph{effective Hamiltonian}.

According to equation \eqref{eq:QFI.pure}, the QFI rate of the state is given by the asymptotic normalised variance of the counting operator
$$
F(\theta) = \lim_{t\to \infty} \frac{4}{t}{\rm Var}(\Lambda_t).
$$
This expression can be interpreted in terms of the theory of \emph{dynamical phase transitions} in open systems \cite{GutaMacieszczakGarrahanLesanovsky}. Borrowing the language of statistical mechanics, one considers counting trajectories as random `configurations', with time playing the role of the extensive variable. Systems near a dynamical phase transitions exhibit trajectories which switch between active (high counting rate) and inactive (low counting rate) `phases' on time scales of the order of the inverse spectral gap of $\mathcal{L}$, and consequently have a large counting variance. This example points to a deeper connection between quantum enhanced metrology and dynamical phase transitions, which is currently under investigation (see also the related paper \cite{Plenio}).   

As in the case of independent ensembles, the  quantum Cram\'{e}r-Rao bound for Markov dynamics is achievable asymptotically \emph{with respect to time} for one-dimensional parameters, while for multidimensional parameters one needs to consider the corresponding Holevo bound for the mean square error. This follows from a general local asymptotic normality result which shows that for large times the output state model can be approximated by a simpler Gaussian shift model \cite{GK17} for which such bounds can be verified directly \cite{GutaGill}. However, standard measurements such as counting and homodyne are in general not optimal, and devising realistic optimal measurements is still an open problem in general.

\subsection{Parameter estimation for standard measurements}
\label{sec:ML}

Traditionally, the study of estimation of dynamical parameters of quantum open systems has focused on the standard classes of continuous-time measurements: counting and homodyne/heterodyne detection. The problem was first posed by Mabuchi 
\cite{Mab96} who considered the estimation of the Rabi frequency of a 
two-level atom in a driven cavity, based on counting trajectories of the photons leaving the cavity. A more refined analysis was carried out in \cite{GW01}, which compared the performance of different measurement schemes, and considered the trade-offs between  estimating of dynamical parameters and the initial system state.

Although both works adopt a Bayesian estimation framework, the problem can equally be posed in the frequentist setting. Suppose that the dynamics $U_\theta$ depends on a parameter $\theta\in \Theta\subset \mathbb{R}^k$, and let $y_{0:t}$ be a continuous time measurement record. As discussed in section \ref{sec:Zakai-likelihood}, the likelihood function of the trajectory is given by  $\theta\mapsto \ell(y_{0:t}|\theta)=
{\rm tr} (\varrho^\theta_t)$ where 
$\varrho^\theta_t$ is the unnormalised conditional state of the principal system, corresponding to the parameter value $\theta$. In numerical implementations it is more convenient to work with the  log-likelihood function which satisfies the equations \eqref{eq:log.lik.h} for for homodyne detection and \eqref{eq:log.lik.c} for counting.
To estimate $\theta$ one can use several likelihood-based methods such as the maximum likelihood estimator (ML) in the frequentist framework,
$$
\hat{\theta}_{\rm ML} = {\arg \max}_{\theta \in \Theta} \log \ell(y_{0:t}|\theta).
$$
and the  posterior mean (PM) and maximum aposteriori (MAP) estimator in Bayesian statistics. To define the latter we will assume that $\theta$ is drawn randomly from a prior distribution $f(d\theta) =f(\theta)d\theta$ over the parameter space $\Theta$. According to the Bayes rule, the \emph{posterior} distribution of $\theta$ is given by:
$$
f(\theta|y_{0:t}) = \frac{\ell(y_{0:T}|\theta)f(\theta)}{\int_{\Theta}\ell(y_{0:T}|\theta^\prime)f(d\theta^\prime )}.
$$
Using the posterior distribution one can construct credible intervals (error bars) and the point estimators mentioned above 
$$
\hat{\theta}_{\rm PM}= \int_\Theta \theta f(\theta|y_{0:t}) d\theta,
\qquad 
\hat{\theta}_{\rm MAP} = {\arg \max}_{\theta \in \Theta} f(\theta|y_{0:t}).
$$

The Bayesian estimation setup has been studied in \cite{GM13}, which derives the equations of the log-likelihood 
function, and provides a Monte-Carlo method for estimating the classical Fisher information of the continuous-measurement. A more detailed analysis of the homodyne measurement case is carried out in \cite{KM16}. Using asymptotic normality results \cite{Burgarth} for time integrated statistics (see also \cite{GutaCatanaBouten,GutavanHorssen}), the authors investigate the classical Fisher information of the integrated homodyne current $W^{Q/P}_{o,t}$  as well as the additional information contained in the two-time correlation statistics 
$$
C^{(2)}_{k,t}= \int_{0\leq u<v\leq t}
k(v-u) dW^{Q/P}_{o,u} dW^{Q/P}_{o,v}
$$
where $k$ is a kernel function. Similarly, for counting measurements one can consider the total number of counts statistic $\Lambda_{o,t}$ \cite{GutaCatanaBouten}; this satisfies the Central Limit Theorem
$$
\frac{1}{\sqrt{t}}(\Lambda_{o,t} - t \mu^\theta) 
\xrightarrow{t\rightarrow\infty} N(0, V^\theta)
$$
where $\mu^\theta = 
{\rm Tr} (\rho^\theta_{ss}L^{\theta\dagger} L^\theta)$ is the counting rate, and the convergence holds in law to the centered, normal distribution with variance
$$
V^\theta ={\rm Tr}\left[ \rho^\theta_{ss} (L^{\theta \dagger}L^\theta + 2 L^{\theta\dagger} A^\theta L^\theta ) 
\right],
$$
where  $A^\theta = \mathcal{L}_\theta^{-1} \left(  L^{\theta\dagger}L^\theta -{\rm Tr}\left[ \rho^\theta_{ss} L^{\theta\dagger}L^\theta \right] I 
\right)$.
From this, one can compute the asymptotic classical Fisher information of the total counts statistics as the Gaussian signal to noise ratio
$$
I_c (\theta) = \frac{\left(\dot{\mu}^\theta\right)^2}{V^\theta},\qquad \dot{\mu}^\theta = \frac{d\mu^\theta}{d\theta}.
$$
Such `linear' statistics are straightforward to compute, and in specific models can be a viable alternative to the general methods described above, as the latter tend to be computationally expensive, especially for multi-dimensional estimation problems. Other Bayesian methods such as approximate Bayesian computation (ABC) \cite{Martin} can be used without the need to explicitly compute the likelihood \cite{GutaCatanaKypraios}. Here the general idea is that the experimental data is compared repeatedly with simulation data (generated for random parameter values), according to certain statistically meaningful distances. The parameters for which the synthetic data is `close' to the real data are retained to build an approximation of the posterior distribution.

So far we have assumed that our measurement data is obtained exclusively by monitoring the environment. However, depending on the experimental setup, it may be possible to perform an additional final measurement $M$ on the system after the trajectory $y_{0:t}$ has been generated. At this point the system's state is given by the filter 
$\rho^\theta_{c,t}$ computed in equations \eqref{eq:state.filter.h}, and \eqref{eq:state.filter.c} and the maximum amount of information that can be extracted is the QFI of the conditional state, denoted $F_s(\theta | y_{0:t})$. The corresponding Cram\'{e}r-Rao bound (for one-dimensional parameters) can be written as 
\cite{ARPG17,GC}
$$
{\rm Var} (\hat{\theta}) \geq \frac{1}{I_{o,t} + \mathbb{E} F_s(\theta | Y_{0:t})}
$$
where $I_{o,t} $ is the classical Fisher information of the output, and the second term is the expected QFI of the conditional system state. The usefulness of the last step depends strongly on the output measurement. If the latter achieves the QFI rate \eqref{eq:qfi.markov} then the final measurement can at most add a sub-linear contribution to the total Fisher information \cite{GK17}. On the other hand, for certain models, the continuous measurement may provide no information while the conditional state does. The scheme was also shown to be useful in achieving Heisenberg scaling 
in a quantum magnetometry problem \cite{ARPG17}.

\section{Quantum system identification}
\label{sec:qsys-id}
Quantum parameter estimation is essentially based on the assumption that one knows in advance everything about the system, for example from first principles modelling, except for the values of a few unknown parameters, which will be determined by a parameter estimation procedure. 

In the problem of quantum system identification for continuously-observed quantum systems, not much is known about the system beyond that it can be described by a QSDE of the form \eqref{eq:HP-QSDE} and that it can be observed continuously. That is, the system is essentially a black box and therefore quantum system identification is in essence a blackbox modeling procedure. 

The aim in quantum system identification is to build and validate a model for the unknown system based on the continuously observed data. As in stochastic modelling and system identification for classical systems \cite{Ljung99}, the procedure typically proceeds in the following stages:
\begin{enumerate}
    \item A candidate model is proposed. The system will have a candidate Hilbert space and is interacting with a traveling field via the QSDE \eqref{eq:HP-QSDE}.
    
    \item Some assumptions are made on the structure of the QSDE, for instance assumptions about the forms of the operators $S$, $L$ and $H$.
    
    \item Continuous-observation data is collected from the system. Part of the record is used for estimating the parameters of the candidate model (Step 4 below) while another part of the record is used to validate the model (Step 5 below)
    
    \item A quantum parameter estimation procedure is developed to estimate the $S$, $L$ and $H$ parameters of the model 
    
    \item Validation of the model is performed to assess if it can adequately model previously unseen observation data
\end{enumerate}

Below we discuss the quantum system identification problem for both non-linear and linear ergodic input output systems.  

\subsection{Quantum system identification of ergodic quantum I/O systems}
\label{subsec:ergodic-IO}
Here we describe the results of the quantum system identification theory for {\em ergodic} quantum I/O systems developed in G{u}\c{t}\u{a} and Kiukas \cite{GK17}.

Consider a finite dimensional,  ergodic quantum I/O system with dynamical parameters 
$D= (H,L)$ and unique, full rank stationary state $\rho_{ss}^D$. 
As we are interested in the long time identification theory, we will assume that the dynamics is stationary, which is equivalent to the principal system starting in the stationary state. Note that for other initial states it may be possible to acquire information about parameters which are not identifiable in the stationary regime. However, it can be shown \cite{GK17} that the associated Fisher information does not scale linearly with time and therefore the parameters cannot be estimated at standard precision scaling.

The stationary output state at time $t$ is 
\begin{align*}
  \rho^{D}_{o,t} = \mathrm{tr}_{\mathcal{H}_{\rm s}}(U^D_t \rho^D_{ss} \otimes |0_{\rm f}\rangle \langle 0_{\rm f}| U_t^{D \dag}),  
\end{align*}
where $\mathcal{H}_{\rm s}$ is the Hilbert space of the principal system. Two I/O systems with parameters $D= (H,L)$ and $D^\prime= (H^\prime, L^\prime)$ are called \emph{output equivalent} if their stationary output states are the same, i.e. $\rho^{D}_{o,t} = \rho^{D^\prime}_{o,t}$, for all times $t$. In \cite{GK17} it was shown that two systems are equivalent if and only if their parameters are related by the action of the cartesian product of a real translation group and the projective unitary group described by the following transformations\footnote{We note that \cite{GK17} treats the more general case of multiple input-output channels}

\begin{itemize}
    \item[(HS)] Shifting the Hamiltonian by a real number $r$ $(H,L)\rightarrow (H+r I,L)$.
    
    \item[(UC)] Unitary conjugation, $(H,L)\rightarrow (W^{\dag}HW,W^{\dag}LW)$ for any unitary operator $W$ on the principal system. 
\end{itemize}
More precisely, the group $G= \mathbb{R}\times U(d)$ acts on the space of parameters 
$$
\mathcal{D}_{erg} = 
\{D= (H,L) : U^{D} {\rm ~is ~ergodic~} \} \subset M_{sa}(\mathbb{C}^d) \times M(\mathbb{C}^D)
$$ 
where $M_{sa}(\mathbb{C}^d)$ denotes the space of selfadjoint matrices, and  the orbits of this action are the equivalence classes of undistinguishable parameters. Along such orbits the QFI rate \eqref{eq:qfi.markov} is equal to zero, which means that phase and unitary conjugation parameters cannot be estimated at standard rate even if the systems were started in a non-stationary state. The \emph{identifiable} parameters are then described by the quotient space of equivalence classes $\mathcal{P}:= \mathcal{D}_{erg}/G$. Thanks to the fact that the group action is free, the quotient has a smooth manifold structure and the QFI rate \eqref{eq:qfi.markov} induces a non-degenerate Riemannian metric on $\mathcal{P}$. 

\subsection{Quantum system identification of linear quantum systems}
\label{subsec:lqs-id} 

Linear quantum systems are a special class of quantum I/O systems that can be viewed as the quantum analogue of continuous-time classical linear stochastic systems; for a detailed introduction, see \cite{NY17}. They model a wide range of linear quantum devices in quantum optics, optomechanics and superconducting circuits, which are of interest for continuous-variable quantum information processing with Gaussian states,  linear quantum signal processing and  sensing. These include devices such as optical and microwave cavities,  parametric amplifiers, linear quantum memories and gravitational wave interferometers; see \cite[Chapters 1 and 6]{NY17} and the references therein.  

A linear quantum system represents a collection of single-mode quantum harmonic oscillators that are mutually coupled by a quadratic Hamiltonian and the oscillators are also linearly coupled to an external  traveling field. Let there be $n$ oscillators with the position and momentum operators collected in the column vector $x=(q_1,p_1,\ldots,q_n,p_n)^T$. The Hamiltonian of the system takes the form $H = \frac{1}{2} x^T R x$, where $R$ is a real symmetric $2n \times 2n$ matrix. 
The coupling operator takes the form $\tilde{L} = K x$ for some complex row vector $K$ of length $n$. The travelling field will be taken to be in a coherent state $|f\rangle$ or more general Gaussian states, where $f$ is a complex square-integrable amplitude function ($\int_{0}^{\infty} |f(\tau)|^2d\tau <\infty$) given by $f(t) = f_R(t) + i f_I(t)$, where $f_R$ and $f_I$ are real-valued functions. 
For simplicity, we will only discuss the case with $S=I$. The time evolution of the vector $x$ is given by the vector $X_t =(j_t(q_1),j_t(p_1),\ldots,j_t(q_n),j_t(p_n))^T$, which satisfies the linear QSDE:
\begin{eqnarray} \label{eq:lqs}
\begin{split}
dX_t &=& A X_tdt + B(\tilde{f}(t)dt + dW_t),\, X_0=x\\
dW_{{ o},t} &=& CX_t dt + \tilde{f}(t)dt + dW_t,\, W_{{o},0}=0,
\end{split} 
\end{eqnarray}
where $\tilde{f}(t)=(f_R(t),f_I(t))^{\top}$, $W_t=(W^{ Q}_t,W^{ P}_t)^{\top}$, $W_{{o},t}=(W^{ Q}_{{\rm o},t},W^{ P}_{{ o},t})^{\top}$. Unlike classical linear stochastic systems, quantum mechanics imposes a constraint on the $A,B,C$ matrices of linear quantum systems, known as the \emph{physical realizability constraints} \cite{NY17}:
\begin{align}
\begin{split}
  A \mathbb{J}_n + \mathbb{J}_n A^{\top} + B \mathbb{J} B^{\top} &=0,\\
\mathbb{J}_nC^{\top} + B\mathbb{J}  &=0,
  \end{split} \label{eq:pr-1}
\end{align}
where $\mathbb{J}  = \left[\begin{array}{cc} 0 & 1 \\ -1 & 0 \end{array}\right]$ and $\mathbb{J}_n  = I_n \otimes \mathbb{J}$.

Consider the case of asymptotically stable linear quantum systems where the $A$ matrix is Hurwitz (all its eigenvalues lie in the left half plane). In the Laplace domain, the input and output fields are related by 
$$
\widetilde{W}_{o,s}= \Xi(s)\widetilde{W}_{s}
$$
where 
$$
\Xi(s)=C(sI-A)^{-1}B + I.
$$
is the \emph{transfer function} and $\widetilde{Y}_s = \int_{0-}^\infty e^{-st} dY_t$ is the Laplace transform of the process $Y$. In an asymptotic setting, this means that if the input is prepared in a state of the frequency mode $\omega$, with mean $m$ then the output will be a of the same mode with mean $m^\prime  =  \Xi(-i\omega) m$. Therefore by probing the systems with time-dependent inputs (e.g. coherent signals) and measuring the corresponding outputs, we obtain information about the transfer function $\Xi(s)$, and implicitly about the system parameters $H,K$. 
The system identification problem in the time-dependent input setting is twofold \cite{GY16,LG17}: firstly to characterise which systems are equivalent, i.e. have the same transfer, and secondly how to estimate the identifiable parameters.

The first question can be answered by appealing to the notion of minimal realisation. For a given transfer function $\Xi(s)$, a linear quantum system $G=(A,B,C)$ is said to be a realization of $\Xi(s)$ if $C(sI-A)^{-1}B+I = \Xi(s)$ and it is said to be a minimal realization if there is no other linear quantum system with fewer oscillators that realize the same transfer function. In \cite{LG17}, generalizing a result of \cite{GY16} for the special case of passive linear quantum systems, it was shown that if two minimal linear quantum systems $G_1=(A_1,B_1,C_1,I)$ and $G_2=(A_2,B_2,C_2,I)$ have the same transfer function $\Xi$ then they have the same dimension and there exists a \emph{symplectic} matrix $V$ such that 
\begin{equation}\label{eq:symplectic-tr}
A_2 = V A_1 V^{-1},  B_2 = V B_1
\quad {\rm and~} C_2 = C_1 V^{-1}.
\end{equation}
This means that the identifiable parameters form the quotient of the space of (stable, minimal) system matrices $G$ by the action \eqref{eq:symplectic-tr} of the symplectic group.

We now consider a second, time-independent system identification setting analysed in \cite{LG17,LGN18}, which is closer in spirit to the non-linear system identification problem of section \ref{subsec:ergodic-IO}. In this case, the input field is prepared in a stationary zero-mean pure Gaussian state (quantum noise) with (symmetrized) covariance matrix $\Gamma$
$$
\frac{1}{2}\langle (dW_t dW_t^{\top} + (dW_t dW_t)^{\top}) \rangle = \Gamma dt.
$$
In this case the output $W_{{o}}$ is a stationary Gaussian processes and is completely characterised by it covariance or \emph{power spectrum}. The SISO case was treated in \cite{LG17} while the general multiple input-multiple output (MIMO) case was given in \cite{LGN18}. 
For a given linear quantum system  $G$, the output power spectrum $\Phi(i\omega)$ of $W_{{ o}}$ is defined as 
$$
\Phi(i\omega ) = \Xi_G(i\omega)^{\#}\, \Gamma\, 
\Xi_G(i\omega)^{\top},\;\omega \in \mathbb{R},\,
$$
where $\Xi_G$ is the transfer function of $G$. For a given input covariance 
$\Gamma$, a linear quantum system $G$ is said to be \emph{globally minimal} if there is no linear quantum system $G'$ with a smaller number of oscillators that has the same power spectrum.
For the class of systems with $D=I$, global minimality is equivalent to minimality, and two globally minimal systems have the same transfer function. In this case it follows that the equivalence class of globally minimum linear quantum systems are those related by a symplectic similarity transformation. That is, if $G_1=(A_1,B_1,C_1,I)$ and  $G_2=(A_2,B_2,C_2,I)$ are globally minimal with the same power spectrum, there is symplectic matrix $V$ such that $A_2 = V A_1 V^{-1}$, $A_2 = V A_1 V^{-1}$ and $C_2=C_1 V^{-1}$.

There is a subtlety when one considers the case where $D$ is symplectic and not a priori fixed to some value (like $D=I$). In this case, two globally minimal systems need not have the same transfer function. To see this, suppose that the input field is in the vacuum state. Then the systems $G=(A,B,C,D)$ and $G'=G=(A,BO,C,DO)$ have the same output power spectrum for any orthogonal-symplectic matrix $O \neq I$ (that is $O$ is both symplectic and orthogonal) but they will not have the same transfer function. Indeed $\Xi_{G'}=\Xi_G O$. To force two globally minimal systems to have the same transfer function we require that the input covariance matrix $\Gamma$ should satisfy the condition that $V \Gamma V^{\top} = \Gamma \Rightarrow V = I$. When this condition on $\Gamma$ is satisfied then again the equivalence class of globally minimal linear quaatum systems with the same output power spectrum is equal to the equivalence class of minimal linear quantum systems with the same transfer function. Note that when $D$ is unknown, the condition imposed on $\Gamma$ implies that the vacuum input should not be used as it does not satisfy this condition. 

Given that the equivalence class of minimal linear quantum systems that have the same transfer function or output power spectrum are those whose system matrices are related by a symplectic similarity transformation, we need to generalize the notion of physical realizability to allow this additional degree of freedom. The physical realizability constraints can be generalized to be (again for the case where $D=I_2$):
\begin{align}
\begin{split}
A Z + Z A^{\top} + B \mathbb{J} B^{\top} &=0,\\
ZC^{\top} + B\mathbb{J}  &=0,
\end{split} \label{eq:pr-2}
\end{align}
for some real invertible skew-symmetric $2n \times 2n$ matrix $Z$. Such a matrix $Z$ can be decomposed as  $Z=V\mathbb{J}_n V^{\top}$ for some symplectic matrix $V$. This means that if a system $G=(A,B,C,I)$ satisfies the generalized physical realizability constraint \eqref{eq:pr-2} then $G'=(V^{-1}AV,V^{-1}B,CV,I)$ satisfies the original physical realizability constraint \eqref{eq:pr-1}. The actual physically meaningful system is $G'$ but if a system $G$ is found that satisfies \eqref{eq:pr-2} then a physical realization of $G$ is given  by $G'$ via the symplectic transformation given above.  The utility of \eqref{eq:pr-2} is that due to the additional degree of freedom $Z$, given a transfer function of a linear quantum system it is easier to first determine a system satisfying \eqref{eq:pr-2} having this transfer function; see \cite{SP12}. This system can then be converted to a physical system satisfying \eqref{eq:pr-1}. 

\subsection{Empirical methods for quantum system identification}

The results discussed in sections \ref{subsec:ergodic-IO} and \ref{subsec:lqs-id} are foundational results for quantum system identification of ergodic quantum I/O systems and linear quantum systems, as they give precise statements about what can be extracted about the black-box model when one knows the output field state of an ergodic quantum I/O system or the transfer function or output power spectrum of a linear quantum system. From a practical perspective, however, one cannot have access to any of the latter ideal quantities. The only information that can be gained about the quantum I/O systems in practice is through performing measurements on the output field, in particular continuous measurements. What this means is that the information required to perform quantum system identification,  whether it is the output field state, transfer function or output power spectrum of a linear quantum system, must be estimated from performing measurements on the quantum system to be identified. Thus empirical methods to construct these estimates from measurement data is crucial for the actual practice of quantum system identification of continuously observed quantum I/O systems.

There are important differences that separate system identification for classical input-output systems and their quantum counterpart. This is primarily due to constraints enforced by quantum mechanics in that not all observables are compatible and can be measured simultaneously. For instance, in \eqref{eq:lqs}, the two components of $W_{o}$ in a linear quantum system cannot be simultaneously measured. On the other hand, if \eqref{eq:lqs} were the equations for a classical linear stochastic system then there is no restriction on simultaenously measuring all elements of $W_{o}$. As a consequence, methods for system identification of classical linear stochastic systems based on data from simultaneous measurements of all accessible outputs cannot be applied to linear quantum systems and new approaches are required. 

A step towards developing empirical methods for quantum I/O systems was proposed in \cite{NAC20} for asymptotically stable linear quantum systems $G=(A,B,C,D)$ when $D$ is known. The work considers linear quantum systems driven by a time-varying coherent input field $|f\rangle$ and information is extracted by measuring the amplitude or phase quadrature of the output field of the system. 

The starting point is that under continuous measurement of one of the quadratures at steady state, the evolution of the conditional expectation of $X_t$ given the observation $Y_{0:t}$ up to time $t$ is given  by the quantum Kalman filtering equation:
\begin{align}
d\hat{X}_t = A \hat{X}_t dt + B\tilde{f}_tdt + L_m (dY_t - C_m \hat{X}_t dt -D_m \tilde{f}_t dt),\label{eq:lqs-Kalman}
\end{align}
where $\hat{X}_t =(\pi(q_1),\pi(p_1),\ldots,\pi(q_n),\pi(p_n))^T$ and $L_m$ is the steady-state quantum Kalman filter gain given by,
\begin{align*}
L_m = Q_m C_m^{\top} + B D_m^{\top},
\end{align*}
where $Q_m=Q_m^{\top} \geq 0$ satisfies the algebraic Riccati equation :
\begin{align*}
AQ_m + Q_m A^{\top} + BB^{\top}-(Q_mC_q^{\top} + B D_q^{\top}) (D_mD_m^{\top})^{-1}(Q_mC_m^{\top} + BD_m^{\top})^{\top}=0.
\end{align*} 
The matrices $C_m$ and $D_m$ are determined from $C$ and $D$ by the type of measurement that is being performed. For measurement of $W^Q_o$, $C_m$ and $D_m$ would be the the first row of $C$ and $D$, respectively, while for measurement of $W^P_o$ they would be the second row of the latter matrices.

Consider the classical linear stochastic system:
\begin{align}
\begin{split} \label{eq:innovation-form}
dz_t &= A z_t dt + B\tilde{f}_t dt + L_m d\nu_t,\\
dY_t &= C_m z_t dt + D_m (\tilde{f}_t dt + d\nu_t), 
\end{split} 
\end{align}
where $\nu_t$ is a standard Wiener process.  The system above is the so-called innovation form for the linear stochastic system,
\begin{align*}
dx_t &= A x_t dt + B(\tilde{f}_t dt + dw_t),\\
dY_t &= C_m z_t dt +  D_m (\tilde{f}_t dt + dw_t),
\end{align*}
where $w_t$ is a standard Wiener process.

Many important system identification algorithms to determine the matrices $A,B,C_m$ of a classical linear systems have been developed for the innovation form \eqref{eq:innovation-form}, including subspace identification algorithms \cite{OD96,Qin06}. Since the innovation form coincides with the steady-state quantum Kalman filter for a corresponding linear quantum system under continuous-measurement of $Y$, an obvious approach to identify the system matrices of a linear quantum system is to exploit existing classical system identification algorithms. However, there are two issues that need to be addressed with this approach:
\begin{enumerate}
\item The system matrices $A$ and $B$ identified by the algorithms will in general not satisfy the physical realizability constraints required of a linear quantum system. The constraints also require that $C_m$ must satisfy $Z C_m^{\top}  = B\mathbb{J}D_m^{\top}$ for some invertible skew-symmetric matrix $Z$. 

\item The algorithm only identifies $C_m$ (corresponding to the choice of measurement) rather than the full matrix $C$. 

\end{enumerate}

The second issue can in fact be resolved as follows. Suppose that $B$ and $C_m$ have been identified such that $Z C_m^{\top} = B\mathbb{J}D_m^{\top}$ for some invertible skew-symmetric matrix $Z$. Then, given that $D$ is known, from the physical realizability constraint $Z C^{\top} = B\mathbb{J}D^{\top}$, the other row of $C$ besides $C_m$, which we denote by $C_{m'}$, can be recovered as $C_{m'} = (Z^{-1} B \mathbb{J}D_{m'}^{\top})^{\top}$. Here $D_{m'}$ denotes the other row of $D$ besides $D_m$. Thus, in fact, resolving issue 1 also resolves issue 2. 
To address issue 1, \cite{NAC20} proposes a two-step procedure:
\begin{enumerate}
\item  Fix  a choice of $n$ (the dimension $n$ is typically not known a priori). Use a classical system identification algorithm to identify system matrices $\hat{A},\hat{B},\hat{C}_m$ of the corresponding dimension from the data. These matrices need not satisfy the physical realizability constraints. 

\item Execute a second optimization algorithm to determine another set of system matrices $\overline{A},\overline{B},\overline{C}_m$ that do satisfy the physical realizability constraints and are also close to the original estimate $\hat{A},\hat{B},\hat{C}_m$ according to some suitable cost function. The optimization algorithm can use $\hat{A},\hat{B},\hat{C}_m$ as the starting point or initial guess. 
\end{enumerate} 

In \cite{NAC20}, the above two-step procedure was considered using subspace identification to compute $\hat{A},\hat{B},\hat{C}_m$ based on the measurement data $y_{0:T}$  (over some fixed finite time interval $[0,T]$). Both output amplitude and phase quadrature measurements were considered.  The elements of $\tilde{f}$ were chosen to be independent pseudo-random binary sequences (PRBS) for persistency of excitation \cite{Ljung99}. The matrices $\overline{A},\overline{B},\overline{C}_m$ were determined by minimizing the cost function $M(A,B,C_m) = \frac{1}{2}(\|A-\hat{A}\|^2 + \|B-\hat{B}\|^2 + \|C_m-\hat{C}_m\|^2 )$ over all triplets of system matrices $(A,B,C_m)$ satisfying the physical realizability constraints. It was shown that this optimization problem can be reformulated as a rank constrained LMI problem, by adopting an approach from \cite{NJP09}. Numerical experiments reported in \cite{NAC20} indicate that this approach can succeed in identifying a physically realizable linear quantum system from the measurement data $y_{0:T}$. However, due to the noise $\nu_t$ in the system, the quality of the approximation depends on the amplitude of the PRBS signal used as input to the system, with higher amplitudes giving a better fitting model according to a normalized mean square error criterion. For the examples considered therein, the approach was also able to select the correct  unknown model order $n$ based on the Akaike final prediction error (FPE) criterion \cite{Ljung99}, when models with different values of $n$ are fitted to the same measurement data. 

\section{Conclusion and outlook}
\label{sec:conclu}

This review covers some of the recent developments in parameter estimation and system identification for quantum input-output systems, with an emphasis on mathematical theory and statistical methodology. A first set of results dealt with statistical properties of the output process at the `quantum level', such as the structure of the space of identifiable parameters and the expression of the QFI \cite{Molmer2014,GK17,GutaCatanaBouten}. These results give general bounds on the estimation accuracy for arbitrary measurements; together with the local asymptotic normality theory of \cite{GK17,GutaCatanaBouten} they indicate that a complete asymptotic theory can be developed similarly to that of state estimation, including a Markovian version of the Holevo bound and the existence of optimal estimators with asymptotically normal errors. A second set of results deals with the likelihood theory for `standard measurements' (counting, homodyne, heterodyne) and the statistical analysis of various estimation methods from simple linear estimators to more the more informative but computationally expensive maximum likelihood \cite{GM13,KM16}. Generally, such measurements are not optimal, and a theoretical understanding of their properties is still lacking. An interesting, and little explored area is that between the `quantum level' and the `standard measurements', for instance understanding the effectiveness of adaptive measurements \cite{GW01}, the use of quantum networks and feedback control techniques \cite{GJ09} to enlarge the class of accessible measurements and improve estimation accuracy. A second direction is towards more realistic models including 
un-monitored channels, non-vacuum inputs, non-Markovian dynamics. A third topic of interest concerns the overlap between dynamical phase transitions and estimation, with potential application in high precision metrology \cite{GutaMacieszczakGarrahanLesanovsky}. 

Empirical methods for construction of a black-box quantum I/O model from continuous-measurement data, including estimation of the model parameters, are crucial for applications involving unknown quantum I/O systems but they are still lacking beyond the preliminary study in \cite{NAC20}. Thus the development of empirical methods for linear and non-linear systems will be an important future research direction in quantum system identification.  

\bibliographystyle{elsarticle-num}
\bibliography{qpe}

\end{document}